\documentclass[final]{svjour2}
\usepackage{graphicx}
\usepackage{rotating}
\usepackage{amssymb}
\usepackage{mathptmx}
\usepackage[numbers]{natbib}
\usepackage{subfig}
\usepackage{sidecap}
\sidecaptionvpos{figure}{t}
\makeatletter
\journalname{Journal of Low Temperature Physics}

\bibpunct{}{}{,}{s}{}{,}

\newcommand\apj{ApJ}
\newcommand\ao{Appl. Optics}

\begin{document}

\newcommand{\hdblarrow}{H\makebox[0.9ex][l]{$\downdownarrows$}-}
\def\fullwidth{0.95}                
\def\halfwidth{0.45}                
\title{Low Noise Titanium Nitride KIDs for SuperSpec: A Millimeter-Wave On-Chip Spectrometer}

\author{S. Hailey-Dunsheath$^1$ \and E. Shirokoff$^2$ \and P. S. Barry$^3$ \and C. M. Bradford$^4$ \and S. Chapman$^5$ \and G. Che$^6$ \and J. Glenn$^7$ \and M. Hollister$^1$ \and A. Kov{\'a}cs$^{8,1}$ \and H. G. LeDuc$^4$ \and P. Mauskopf$^{6,3}$ \and C. McKenney$^1$ \and R. O'Brient$^4$ \and S. Padin$^1$ \and T. Reck$^4$ \and C. Shiu$^1$ \and C. E. Tucker$^3$ \and J. Wheeler$^7$ \and R. Williamson$^4$ \and J. Zmuidzinas$^1$}
\institute{1: California Institute of Technology, Mail Code 301-17, 1200 E. California Blvd., Pasadena, CA 91125, USA; 
\email{haileyds@caltech.edu} \\
2: Department of Astronomy \& Astrophysics, University of Chicago, 5640 South Ellis
Avenue, Chicago, IL 60637, USA \\
3: School of Physics \& Astronomy, Cardiff University, 5 The Parade, Cardiff, CF24 3AA, UK \\
4: Jet Propulsion Laboratory, 4800 Oak Grove Drive, Pasadena, CA 91109, USA \\
5: Department of Physics and Atmospheric Science, Dalhousie University, Coburg Road, Halifax, NS B3H 1A6, Canada \\
6: School of Earth and Space Exploration and Department of Physics, Arizona State University, Tempe, AZ 85287, USA \\
7: Center for Astrophysics and Space Astronomy, University of Colorado, 1255 38$^\mathrm{th}$ Street, Boulder, CO 80303, USA \\
8: Institute for Astrophysics, University of Minnesota, 116 Church St SE, Minneapolis, MN 55455, USA}

\maketitle

\begin{abstract}

SuperSpec is a novel on-chip spectrometer we are developing for multi-object, moderate resolution ($R = 100 - 500$), large bandwidth ($\sim$$1.65$:$1$) submillimeter and millimeter survey spectroscopy of high-redshift galaxies. The spectrometer employs a filter bank architecture, and consists of a series of half-wave resonators formed by lithographically-patterned superconducting transmission lines. The signal power admitted by each resonator is detected by a lumped element titanium nitride (TiN) kinetic inductance detector (KID) operating at $100-200$ MHz. We have tested a new prototype device that achieves the targeted $R=100$ resolving power, and has better detector sensitivity and optical efficiency than previous devices. We employ a new method for measuring photon noise using both coherent and thermal sources of radiation to cleanly separate the contributions of shot and wave noise. We report an upper limit to the detector NEP of $1.4\times10^{-17}$ W\,Hz$^{-1/2}$, within 10\% of the photon noise limited NEP for a ground-based $R=100$ spectrometer.


\end{abstract}


\section{Introduction}

The epoch of reionization and the birth and subsequent growth of galaxies in the first half of the Universe's history ($z \gtrsim 1$) are key topics in modern astrophysics. Measurements of the cosmic far-IR background indicate that in aggregate, much if not most of the energy released by stars and accreting black holes over cosmic time has been absorbed and reradiated by dust~\cite{Fixsen1998}. A complete understanding of galaxy evolution since reionization therefore requires observations at (sub)millimeter wavelengths, where the dust emission peaks, and the extinction of diagnostic spectral lines is minimized. Survey spectroscopy at (sub)millimeter wavelengths, using a multi-beam spectrometer such as we describe here, is uniquely poised to access the high-redshift Universe, both through the measurement of individual galaxies, and via statistical studies in wide-field tomography~\cite{Crites2014SPIE}. In particular, the 158 $\mu$m [CII] transition is typically the brightest spectral feature in dusty galaxies, and promises to be a powerful probe of galaxies at redshifts $z \ge 3$, where it is shifted into the telluric windows at $\lambda \ge 600$ $\mu$m~\cite{Wang2013}.

SuperSpec is a novel, ultra-compact spectrometer-on-a-chip for (sub)millimeter wavelength astronomy~\cite{HaileyDunsheath2014SPIE}. Its very small size, wide spectral bandwidth, and highly multiplexed detector readout will enable construction of powerful multi-beam spectrometers for high-redshift observations. We are currently developing this technology with $R \approx 100-500$ prototypes operating in the $190-310$ GHz band. Here we present the design and characterization of a new SuperSpec device that succeeds in achieving a lower resolving power ($R\approx100$), improved detector sensitivity, and higher optical efficiency than previous devices.


\section{Third Generation SuperSpec Prototype}

The SuperSpec chip is designed to be a compact, superconducting filter bank spectrometer~\cite{Kovacs2012SPIE}. Radiation propagating down a transmission line encounters a series of tuned resonant filters, each of which consists of a section of transmission line of length $\lambda_i/2$, where $\lambda_i$ is the resonant wavelength of channel $i$. These half-wave resonators are coupled to the feedline and to power detectors with adjustable coupling strengths, described by quality factors $Q_\mathrm{feed}$ and $Q_\mathrm{det}$, respectively. Accounting for additional sources of dissipation in the circuit with a coupling factor $Q_\mathrm{loss}$, the spectrometer resolving power $R$ is equal to the net filter quality factor $Q_\mathrm{filt}$, and is given by $R^{-1} = Q_\mathrm{filt}^{-1} = Q_\mathrm{feed}^{-1} + Q_\mathrm{det}^{-1} + Q_\mathrm{loss}^{-1}$. The filter bank is formed by arranging a series of channels monotonically decreasing in frequency, with a spacing between channels equal to an odd multiple of $\lambda_i/4$. 

The SuperSpec concept is implemented with thin-film superconducting circuits, and is described more fully elsewhere~\cite{HaileyDunsheath2014SPIE, Barry2012SPIE, Shirokoff2012SPIE, HaileyDunsheath2014LTD, Shirokoff2014LTD}. Free space radiation is coupled into the feedline from a broadband antenna. Both the feedline and the resonators are inverted microstrip, consisting of Nb traces on Si substrate beneath a SiN$_x$ dielectric and Nb ground plane. The signal power admitted by each resonator is dissipated in a segment of lossy meander formed from titanium nitride (TiN)\cite{LeDuc2010}, which is connected in parallel to an interdigitated capacitor (IDC) made from the same TiN film to form a lumped element kinetic inductance detector (KID). Each KID is coupled to a coplanar waveguide readout feedline (CPW) by a small coupling capacitor, formed by TiN on the KID side, and Nb on the readout side. The KIDs have low readout frequencies ($100-200$ MHz) and are high$-Q$ ($Q\sim10^5$) in order minimize the readout bandwidth per channel, thereby maximizing the multiplexing density. The chip is cooled by a $^3$He sorption refrigerator to 220 mK.

Previous SuperSpec prototypes have demonstrated the basic functionality of the filter bank, with spectral channels operating at $R\gtrsim250$~\cite{HaileyDunsheath2014SPIE, HaileyDunsheath2014LTD, Shirokoff2014LTD}. We have fabricated a third generation device with adjustments to the spectrometer layout intended to achieve lower $R\sim100$, which is a natural choice for a tomographic mapping experiment~\cite{Crites2014SPIE}. We have additionally reduced the inductor volume to increase the detector responsivity.

\begin{figure}
\begin{center}
\includegraphics[%
  width=\fullwidth\linewidth,
  keepaspectratio]{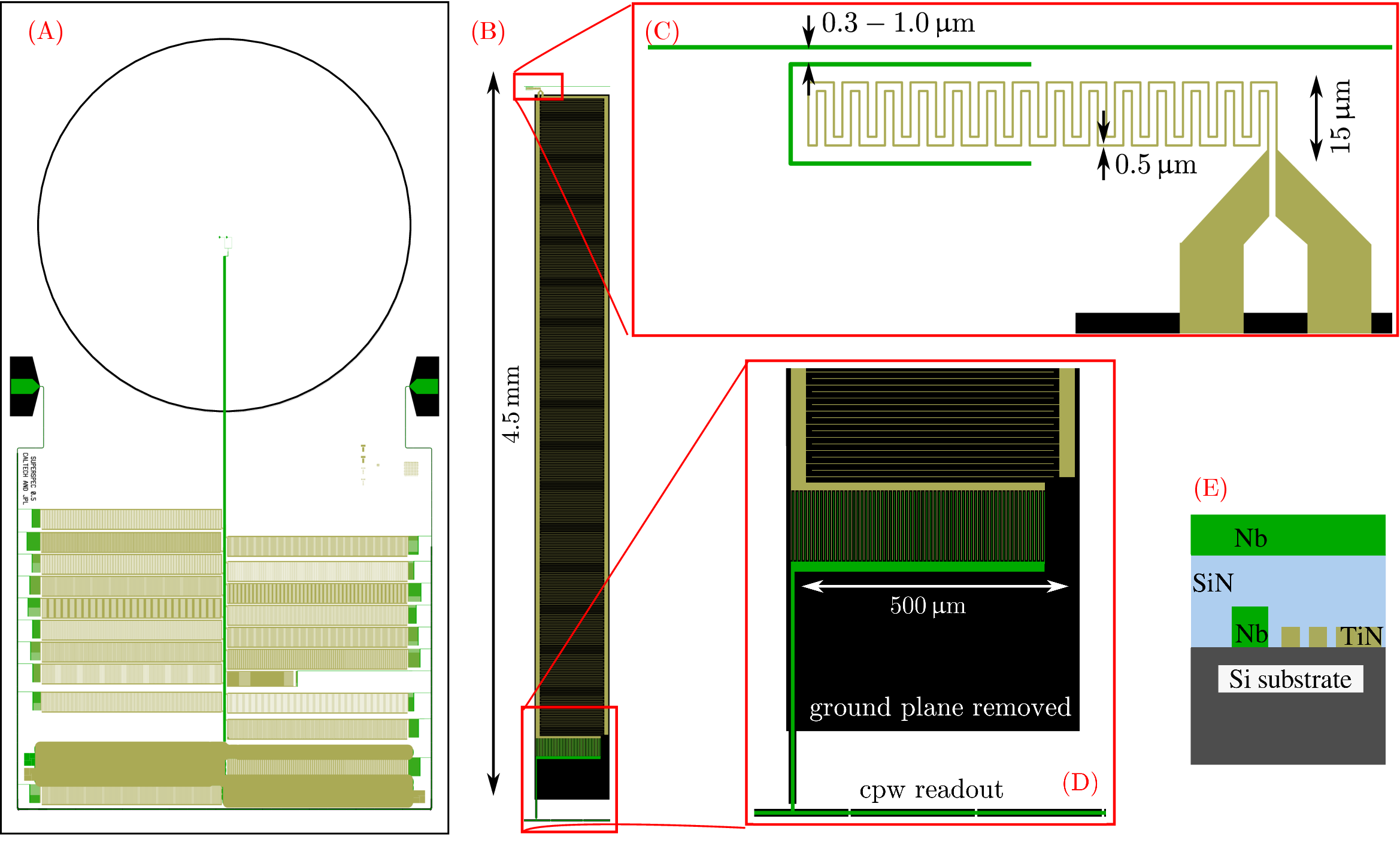}
\end{center}
\caption{\small Gen 3 device design. (\textbf{A}) A test die mask, with dual-slot antenna and lens footprint at the top, and feedline running vertically past an array of filters to a broadband terminator. (\textbf{B}) A single millimeter-wave filter and KID. (\textbf{C}) The millimeter-wave resonator and inductor meander at the top of the KID. (\textbf{D}) The lower portion of the large IDC, coupling capacitor, and readout CPW. (E) Cross-section showing the device layers; in the region surrounding the IDC, SiN$_x$ and ground plane are removed. (Color figure online.)}
\label{fig:newdesign}
\end{figure}

Figure \ref{fig:newdesign} shows the new design. The lower resolving power is achieved by reducing the gap between the Nb feedline and the Nb filter resonator element. The test die includes a bank of 9 spectral channels with gaps varying from $0.3-1.0$ $\mu$m, along with 2 moderate-$Q$ ($Q_\mathrm{filt}\sim300$) channels inherited from previous designs. The increased detector responsivity is achieved in part by reducing the TiN linewidths from $1.0$ to $0.5$ $\mu$m, which reduces the inductor volume to 9 $\mu$m$^3$, a factor of 4 smaller than in the previous generation devices~\cite{HaileyDunsheath2014SPIE}. We have also reduced the TiN $T_c$ from 1.65 K to 1.25 K, which further increases the responsivity~\cite{LeDuc2010}.


The test die also contains two high-$Q$ ($Q_\mathrm{filt} > 1000$) filter channels and an absorber-free millimeter-wave resonator to measure the millimeter-wave loss. Three pairs of broad band absorbers placed directly before the filter channels, directly after the filter channels, and along the terminator enable the measurement of the total power on the line. A dark KID is included to monitor sensitivity to direct stimulation. Coupling to free space radiation is achieved with a dual-slot antenna behind a hyperhemispherical alumina lens, which has an epoxy-based anti-reflection coating~\cite{Rosen2013}.



\section{Spectral Profiles}

\begin{SCfigure}
\centering
\includegraphics[%
  width=0.60\linewidth,
  keepaspectratio]{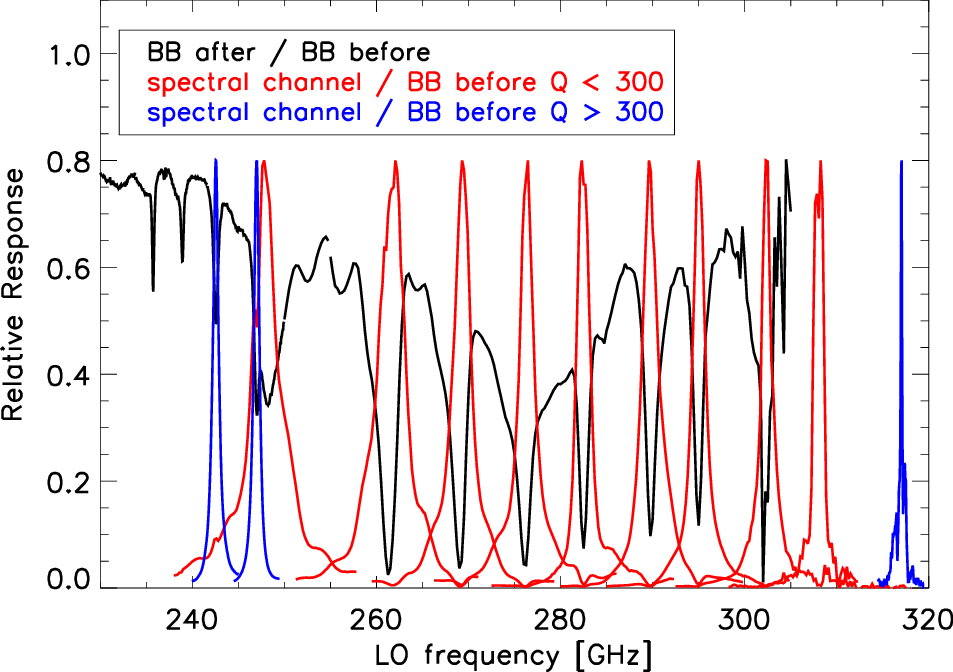}
\caption{\small Ratio of broadband channels after and before the filter bank (\textit{black}), spectral profiles ratioed to a broadband channel before the filter bank for the 9 low-$Q$ channels (\textit{red}), and 3 $Q_\mathrm{filt}>300$ channels (\textit{blue}), with arbitrary scaling. Narrow features in the broadband channel ratio at 235.7 and 238.9 GHz are due to the absorber-free resonator, and a moderate-$Q$ channel missing in readout, respectively. Low-pass metal mesh filters reduce the system transmission above 300 GHz. (Color figure online.)}
\label{fig:global_sockout}
\end{SCfigure}

We probe the optical response of the circuit by making spectral scans with a swept coherent source~\cite{HaileyDunsheath2014LTD}. A ratio of the response of a spectral channel to that of a broadband channel produces the normalized spectral profile, while the relative responses of broadband channels placed before and after the filter bank track the power removed from the line by the resonator. In Figure~\ref{fig:global_sockout} we show the results of a spectral scan of the full circuit. All 9 of the low-$Q$, small gap spectral channels are present. The measured resolving power of these channels is $R\approx100-250$, and increases with increasing gap, as anticipated.


\begin{figure}%
    \centering
    \subfloat
      {{\includegraphics[width=\halfwidth\linewidth]{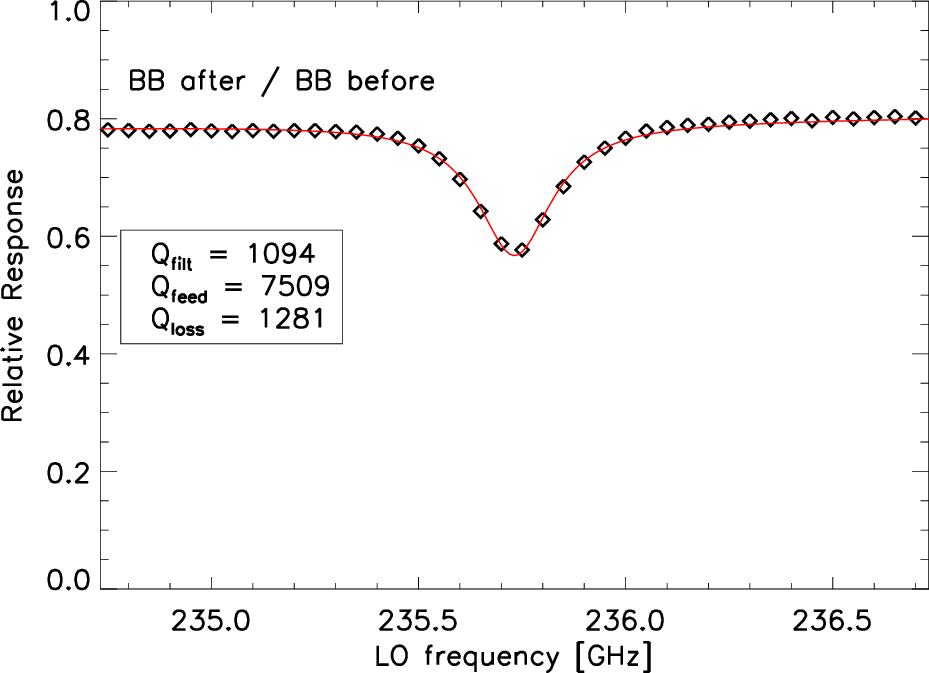}}
      \label{fig:qloss_sockout}}%
    \qquad
    \subfloat
      {{\includegraphics[width=\halfwidth\linewidth]{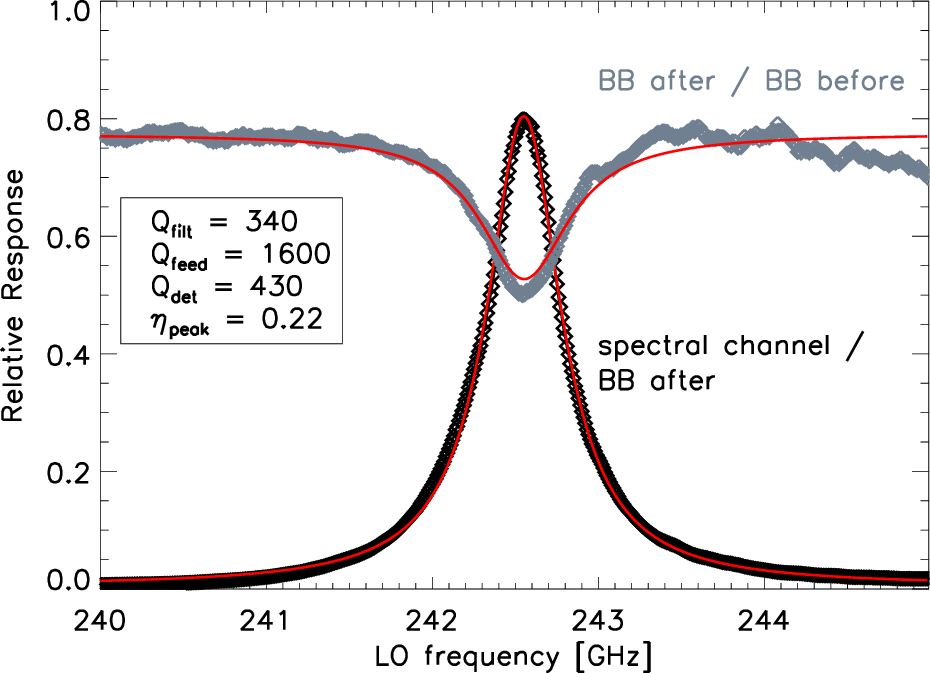}}
      \label{fig:spec_sockout}}%
    \caption{\small (\textbf{a}) Relative response of a pair of broadband channels placed after and before the filter bank, near the resonant frequency of an absorber-free millimeter-wave resonator (\textit{open symbols}). Model fit indicates $Q_\mathrm{loss}=1281$ (\textit{red}). (\textbf{b}) Spectral profile of an $R=340$ channel normalized to a broadband channel placed after the filter bank, with an additional arbitrary scaling (\textit{black}), relative response of a pair of broadband channels placed after and before the filter bank (\textit{gray}), and model fits (\textit{red}). The broadband channel ratio at $\nu \gtrsim 244$ GHz is reduced by a spectral channel at $\nu=248$ GHz. (Color figure online.)}
    \label{fig:sockouts}%
\end{figure}

The broadband channel response ratio shows a narrow minimum at 235.7 GHz, evidence of power removed from the feedline by the absorber-free millimeter-wave resonator (Figure~\ref{fig:qloss_sockout}). The shallow depth of this feature indicates that the bulk of this loss is due to dissipation, with a loss characterized by $Q_\mathrm{loss}=1281$. This dissipation is consistent with previous estimates of loss in the SuperSpec resonators, and is consistent with expectations for loss due to the SiN$_x$ dielectric~\cite{HaileyDunsheath2014LTD}. This loss will reduce the efficiency of each spectral channel by a factor of $\eta_\mathrm{loss} = [1-Q_\mathrm{filt}/Q_\mathrm{loss}]^2$, with $\eta_\mathrm{loss} = 0.59\rightarrow0.85$ for $Q_\mathrm{filt} = 300\rightarrow100$.

The moderate-$Q$ channel at 242.5 GHz has the most well-isolated spectral profile, and we have performed a detailed characterization of the spectral response and sensitivity of this channel. In Figure~\ref{fig:spec_sockout} we show the normalized spectral response and broadband channel ratio. A joint fit to these profiles yields $Q_\mathrm{filt} = 340$. The channel is undercoupled ($Q_\mathrm{det} < Q_\mathrm{feed}$), which reduces the detection efficiency. The combination of undercoupling and additional loss described by $Q_\mathrm{loss}$ results in a modeled on-resonance detection efficiency of 22\%.


\section{Noise Characterization}

\begin{figure}%
    \centering
    \subfloat%
      {{\includegraphics[width=\halfwidth\linewidth]{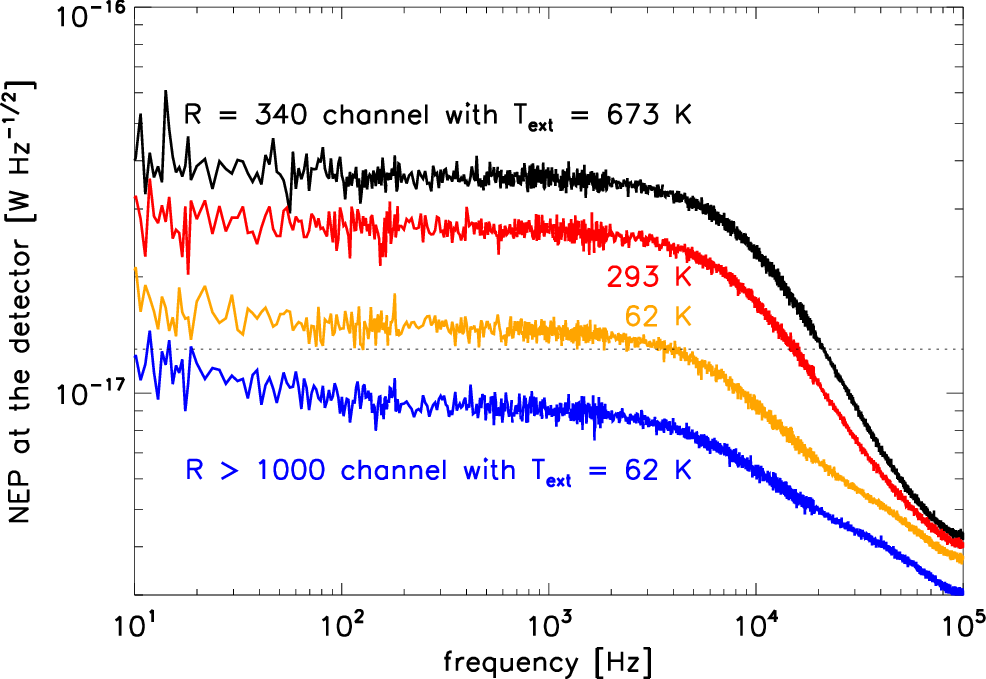}}
      \label{fig:psds_nep}}%
    \qquad
    \subfloat%
      {{\includegraphics[width=\halfwidth\linewidth]{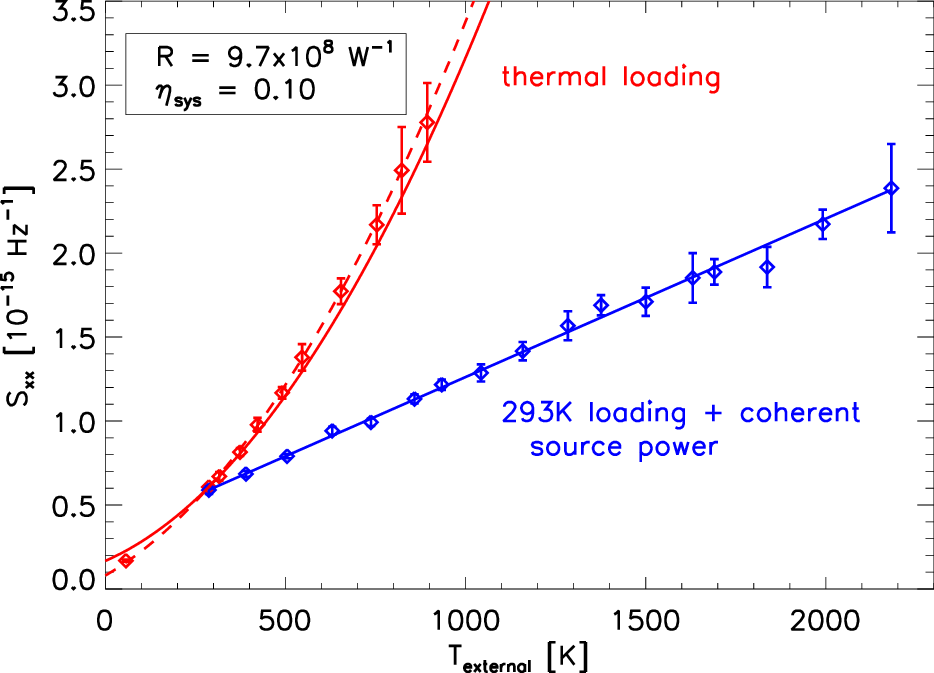}}
      \label{fig:noise_v_loading}}%
    \caption{\small (\textbf{a}) Noise spectra from an $R=340$ channel under external optical loadings of 673 K (\textit{black}), 293 K (\textit{red}), and 62 K (\textit{orange}), referenced to the detector. Also shown is the fractional frequency noise spectrum on a high-$R$ spectral channel (\textit{blue}), converted to an NEP assuming a constant responsivity for all detectors on the die. The background-limited NEP for an $R=100$ ground-based spectrometer is also indicated (\textit{dotted line}). (\textbf{b}) Fractional frequency noise averaged over $400-800$ Hz as a function of loading from a coherent source (\textit{blue points}) and thermal sources (\textit{red points}), along with model fits (\textit{lines}). Models for thermal loading include additional noise due either to a constant device noise (\textit{solid}) or 8 K of internal loading (\textit{dashed}). (Color figure online.)}%
    \label{fig:noise}%
\end{figure}

In Figure~\ref{fig:psds_nep} we show noise spectra measured for the $R=340$ channel under a range of optical loadings. Hot loads ($T_\mathrm{ext} > 273$ K) are achieved by reimaging the beam waist onto the aperture of a cavity blackbody, while a cold load is obtained by placing a mirror in front of the cryostat entrance window. The detector is photon noise limited down to the lowest measured loading, corresponding to $T_\mathrm{ext} = 62$ K.

The scaling of shot noise with optical loading is a standard method of estimating the responsivity and optical efficiency of KIDs~\cite{Yates2011}. Previous measurements have characterized KIDs illuminated by cryogenic blackbody loads, with a high ratio of optical frequency to load temperature ensuring a low occupation number and negligible wave noise~\cite{Yates2011, deVisser2014NatCo, McCarrick2014, Hubmayr2015TiNphotonnoise}. Our approach is to measure the photon noise under loading from a coherent source, which generates pure shot noise. Swapping the coherent source for a hot thermal source, while conserving the detected power, then allows a separate measurement of the wave noise contribution.

In Figure~\ref{fig:noise_v_loading} we show the fractional frequency ($x=\delta f_r/f_r$) noise PSD ($S_{xx}$) with the system exposed to power from either an external thermal load, or from a coherent source. With the coherent source in place the change in $S_{xx}$ with respect to a change in the absorbed optical power $P$ is:

\begin{equation} \label{eq:shot_noise}
\frac{\partial S_{xx}}{\partial P} = \mathcal{R}^2(2h\nu)\bigg[1 + \frac{NEP_\mathrm{rec}^2}{NEP_\mathrm{shot}^2}\bigg],
\end{equation}

\noindent where $\mathcal{R}=dx/dP$ is the fractional frequency responsivity, and the ratio $NEP_\mathrm{rec}^2 / NEP_\mathrm{shot}^2$ accounts for the additional contribution of quasiparticle recombination noise. This ratio may be written as $NEP_\mathrm{rec}^2 / NEP_\mathrm{shot}^2 = 2\Delta/h\nu \eta_s$, where $\Delta$ is the superconducting gap, and $\eta_s$ is the pair-breaking efficiency. For $\nu = 242.5$ GHz we have $h\nu/\Delta \approx 5.3$, and the breaking of Cooper pairs by phonons with energies greater than $2\Delta$ may potentially increase $\eta_s$~\cite{Guruswamy2014}. However, as our films are thin ($t=20$ nm), we assume the time scale for phonon loss from the film is much shorter than the pair-breaking timescale, and set $\eta_s = 2\Delta/h\nu$~\cite{Guruswamy2014}. Equation~\ref{eq:shot_noise} then reduces to $\partial S_{xx} / \partial P = 2\mathcal{R}^2(2h\nu)$.

The scaling of $S_{xx}$ with coherent source power shown in Figure~\ref{fig:noise_v_loading} yields $\mathcal{R} = 9.7\times10^8$ W$^{-1}$, and a full system optical efficiency of $\eta_\mathrm{sys} = 10\%$. This responsivity is significantly higher than in previous SuperSpec devices, and follows the reduction in inductor volume and $T_c$ incorporated in the Gen 3 design. The combined optical efficiency of the filter stack, antenna, and spectrometer is expected to be 13\% for this channel, slightly higher than measured. An improved spectrometer efficiency will result from better matching $Q_\mathrm{feed}$ and $Q_\mathrm{det}$ in future devices. Increasing $\eta_\mathrm{sys}$ to $20\%$ would reduce the background-limited system NEP by $\approx$$25\%$.

The measured $S_{xx}$ under thermal loading is larger than under loading from the coherent source, due to the additional wave noise. For the values of $\mathcal{R}$ and $\eta_\mathrm{sys}$ determined above, the measured noise is in excess of a model that only considers photon noise from the external load. In Figure~\ref{fig:noise_v_loading} we overplot models that account for this excess with either a fixed device noise, or a fixed 8K loading. The latter model is slightly favored, particularly for the point at $T_\mathrm{ext}=62$ K, and provides a constraint on loading from within the cryostat.

Figure~\ref{fig:psds_nep} shows that the noise is still photon-dominated at the lowest optical loading achieved here. This provides an upper limit to the detector-limited NEP of $1.4\times10^{-17}$ W\,Hz$^{-1/2}$, very close to the background-limited NEP value of $1.3\times10^{-17}$ W\,Hz$^{-1/2}$ appropriate for a ground-based $R=100$ spectrometer. We also show the noise measured for a narrower bandwidth $R > 1000$ spectral channel ($\nu=317$ GHz; Figure~\ref{fig:global_sockout}), in which the fractional frequency noise $S_{xx}$ has been converted to an NEP $<10^{-17}$ W\,Hz$^{-1/2}$ assuming the same responsivity as measured for the $R=340$ channel. This narrow band channel has a low optical loading, and the low photon-limited NEP inferred here suggests the current detectors are sufficiently sensitive for background-limited spectroscopy at $R=100$. Future dark tests will provide robust measurements of the detector NEP. 

\section{Summary}

We present the design and characterization of a new SuperSpec prototype device. By reducing the gap between the feedline and filter resonator element to 0.3 $\mu$m we have succeeded in achieving resolving powers as low as $R\approx100$, ideal for a tomographic mapping experiment. Reductions in the TiN linewidth and $T_c$ have increased the detector responsivity. We characterize a single spectral channel using a new method to separately measure the shot and wave noise from a thermal source. We find a full system optical efficiency of $\eta_\mathrm{sys} = 10\%$ and an upper limit to the detector NEP of $1.4\times10^{-17}$ W\,Hz$^{-1/2}$, within $10\%$ of the targeted photon noise-limited value. 

\begin{acknowledgements}
This project is supported by NASA Astrophysics Research and Analysis (APRA) grant no. 399131.02.06.03.43, and by the NSF Advanced Technology and Instrumentation program award \#1407287. 
\end{acknowledgements}


\end{document}